\PassOptionsToPackage{caption=false,font=footnotesize}{subfig}
\documentclass[journal]{IEEEtran}
\usepackage{amsthm}
\usepackage{amsfonts}
\usepackage[normalem]{ulem}
\usepackage{amssymb}
\usepackage{subfig}
\usepackage[top=0.67in, bottom=0.88in, left=0.625in, right=0.625in]{geometry}
\ifCLASSINFOpdf
  \usepackage[pdftex]{graphicx}
  \usepackage{graphicx}
  \graphicspath{{../pdf/}{../jpeg/}{../png/}}
  \DeclareGraphicsExtensions{.pdf,.jpeg,.png}
\else
  \usepackage[dvips]{graphicx}
  \graphicspath{{../eps/}}
  \DeclareGraphicsExtensions{.eps}
\fi
\ifCLASSOPTIONcompsoc
  \usepackage[caption=false,font=normalsize,labelfont=sf,textfont=sf]{subfig}
\else
\usepackage[caption=false,font=footnotesize]{subfig}
\fi
\usepackage{multirow}
\usepackage{balance}
\usepackage{multicol}
\usepackage{epstopdf}
\usepackage{verbatim} 

\usepackage[english]{babel}
\usepackage{float}
\usepackage{xcolor}
\usepackage[linesnumbered,ruled,vlined]{algorithm2e}

\usepackage[hidelinks]{hyperref}
\usepackage{cite}
\renewcommand{\footnoterule}{%
  \kern -3pt 
  \hrule width 1\columnwidth height 0.4pt 
  \kern 2.6pt 
}
\ifCLASSINFOpdf
   \usepackage[pdftex]{graphicx}
   \else
\usepackage[dvips]{graphicx}
\fi
\usepackage[cmex10]{amsmath}
\usepackage[T1]{fontenc}
\newcommand*{\affmark}[1][*]{\textsuperscript{\dag}}

\begin{document}
\title{SEMIKHORN: Globally balanced affinities for mmWave Localization in MU mMIMO systems \vspace{-0.4 \baselineskip}}
\author{
\normalsize{Abhisha~Garg~\IEEEmembership{Graduate Student Member,~IEEE,} }}
\author{
\normalsize{Abhisha~Garg$^{*}$, Raghav~Shukla$^{*}$, Suraj~Srivastava$^{\dagger}$, Aditya~K.~Jagannatham$^{*}$,\\ Department of Electrical Engineering, Indian Institute of Technology Kanpur, India$^*$ \\ Department of Electrical Engineering, Indian Institute of Technology Jodhpur, India$^\dagger$\\(e-mail: abhisha20@iitk.ac.in$^*$; raghavs21@iitk.ac.in$^*$, surajsri@iitj.ac.in$^\dagger$; adityaj@iitk.ac.in$^*$)\vspace{-2.5 \baselineskip}}}
\maketitle
\thispagestyle{empty} 
\pagestyle{empty} 
\begin{abstract}
This work conceives \textit{SEMIKHORN}, a semi-supervised channel charting (CC) framework for mmWave localization, which leverages t-SNEkhorn, a doubly stochastic variant of t-distributed Stochastic Neighbor Embedding (t-SNE) that utilizes entropic optimal transport to construct pairwise similarities. Unlike standard t-SNE, which normalizes affinities independently for each data point, t-SNEkhorn generates globally balanced similarities ensuring consistent neighborhood representation. We consider wireless networks with distributed base stations (BSs) equipped with multiple antennas, where each BS constructs a local dissimilarity matrix from the channel state information (CSI). These local dissimilarity matrices are then fused to obtain a single global dissimilarity matrix, which is processed through manifold learning to embed users onto a geometric map. The performance is evaluated in a simulated outdoor environment, and Bayesian optimization is employed on the framework hyperparameters to minimize the mean localization error (MLE). Experimental results demonstrate that the proposed framework achieves an MLE of $6.86\%$ in a circular vicinity of radius $100$m, requiring less than $15\%$ of labeled CSI samples.
\end{abstract}
\vspace{-1mm}
\begin{IEEEkeywords}
Channel charting, manifold learning, affinity, t-distributed stochastic neighbor embedding, Bayesian optimization
\end{IEEEkeywords}
\IEEEpeerreviewmaketitle
\vspace{-7mm}
\section{Introduction} \label{intro}
\vspace{-1mm}
mmWave band, spanning from $30-100$ GHz \cite{heath2016overview}, is emerging as a promising candidate for future wireless systems due to its ideal balance between bandwidth availability and moderate propagation losses. However, accurate user localization becomes a critical prerequisite for harnessing the full potential of mmWave systems. For many location-based services, absolute user positions are not strictly required; instead, pseudo-locations that preserve local neighborhood relationships are often sufficient \cite{deng2018multipoint}. In this regard, Channel Charting (CC) \cite{studer2018channel} has recently emerged as a promising alternative. Studer \textit{et al.}, in their ground-breaking work, proposed a novel framework that learns a latent representation of the radio geometry, effectively capturing relative user locations and spatial dynamics without requiring ground-truth position data. However, the existing framework relies on a single BS, which can only capture limited spatial relationships. As a result, users located in different cells may exhibit distorted neighborhood relationships in the learned chart. To overcome these limitations, multi-point CC \cite{deng2018multipoint} leverages CSI collected from multiple geographically distributed BSs, enabling a more comprehensive and spatially consistent representation of UE locations. However, the framework only preserves relative neighborhood relationships among users and still lacks absolute spatial information. To address this limitation, semi-supervised learning can be incorporated by introducing a small set of labeled user positions as geometric anchors. Wu \textit{et al.} \cite{wu2025improved}, proposed a deep learning-based triplet channel charting algorithm to infer absolute geographical positions. They introduced a covariance-based channel feature to improve inference performance under low-SNR conditions. Zhang and Saad \cite{zhang2021semi} proposed a convolutional autoencoder-based model that employs semi-supervised CC to infer the true UE locations from channel features. Moreover, while machine learning-based approaches can achieve high localization accuracy through powerful \textit{data-driven} representations, they typically require extensive labeled data, careful hyperparameter tuning, and substantial computational resources, which limits their feasibility and scalability. Deng \textit{et al.} \cite{deng2021semi}, proposed a t-distributed Stochastic Neighbor Embedding (t-SNE) based semi-supervised localization technique to map the CSI data to low-dimensional embedding. However, standard t-SNE performs asymmetric affinity normalization, which can distort the global geometry and lead to unbalanced neighborhood representations in the embedded space\footnote{Note, affinity \& similarity are used interchangeably throughout the work.}. To address these drawbacks, we construct doubly stochastic affinities tailored to wireless channel characteristics using the Sinkhorn-Knopp \cite{van2023snekhorn} algorithm. These improved affinities are then incorporated into the t-SNE objective, giving rise to t-SNEkhorn, the foundation upon which we build our semi-supervised framework, \textit{SEMIKHORN}. The contributions are summarized next.
\vspace{-2mm}
\subsection{Contributions} \label{contri}
\vspace{-1mm}
We propose the first application of symmetric and doubly stochastic entropic affinities\footnote{Entropic affinities refer to the similarity matrix generated through entropic optimal transport as detailed in Section-\ref{entropic_OT}.}, via t-SNEkhorn, for semi-supervised wireless localization replacing the Euclidean projection onto the space of symmetric matrices used in t-SNE. t-SNEkhorn enforces global symmetry while preserving per-point entropy, thereby producing more balanced neighborhood graphs for CC. From a theoretical standpoint, we analytically show that non-doubly stochastic affinity matrices in t-SNE introduce a positive lower bound on the Kullback–Leibler (KL) divergence objective, which limits the optimization of the cost function. However, enforcing double stochasticity in t-SNEkhorn relaxes this constraint by removing the non-zero lower bound, enabling more effective minimization of the KL objective and improving embedding fidelity. Furthermore, the doubly stochastic constraint aligns naturally with the uniformly distributed UEs considered in this work, as it enforces balanced pairwise similarities and uniform perplexity, ensuring that spatially uniform users form uniformly connected neighborhood graphs. We further reformulate the dissimilarity measure using the Manhattan (L1) metric, which provides a \textit{stronger correlation} between the channel features and spatial proximity compared to the conventional Euclidean (L2) metric. Lastly, \textit{Bayesian optimization} is considered over manual tuning to compute optimal embedding hyperparameters, which yields substantial gains in embedding trustworthiness and reduces localization error.
\vspace{-2mm}
\subsection{Notations:} The quantity $\langle \cdot, \cdot \rangle$ represents inner product, $\parallel \! \cdot \! \parallel_{\mathcal{F}}$ denotes Frobenius norm, $m_r(\cdot)$ and $m_c(\cdot)$ denote row-sum and column-sum operation, respectively, while $\oplus$ represents the outer sum. The element-wise exponential (Hadamard exponential) is denoted as $\mathrm{exp}^\odot(\cdot)$, while Hadamard division is given by $\oslash$. $[x]_+ = \mathrm{max}(x,0)$, L1 and L-infinity norm is represented as $\parallel\cdot\parallel_1$ and $\parallel\cdot\parallel_\infty$, while $\min(\cdot)$ and $\mathrm{\max(\cdot)}$ represent the minimum and maximum values of a function respectively.
\vspace{-2mm}
\section{Channel Model and mmWave localization} \label{sys_model}
\vspace{-1mm}
Consider a mmWave massive MIMO (mMIMO) system where $B$ base stations, each equipped with $N$ antennas serving $K$ single-antenna UEs. Each BS employs a uniform linear array (ULA), and the channel between the $k$-th user and the $b$-th BS within one coherence bandwidth can be modeled as
\vspace{-1mm}
\begin{align}
    \mathbf{h}_{k}^b = \sum_{l=1}^{L_{k}^b}\gamma_{k}^{b,l}\mathbf{a}(\theta_{k}^{b,l}),
\end{align}
where $L_{k}^b$ represents the number of multipath components, $\theta_{k}^{b,l}$ denotes the angle-of-arrival (AoA) for the $l$-th path and $\gamma_{k}^{b,l}$ represents the complex path gain corresponding to the $l$-th path. The array steering vector $\mathbf{a}(\theta) \in \mathbb{C}^{N \times 1}$ is defined as
\vspace{-1mm}
\begin{align}
    \mathbf{a}(\mathbf{\theta}) = \frac{1}{\sqrt{N}}\big[1, e^{-j\frac{2\pi}{\lambda}d \cos\theta}, \cdots, e^{-j\frac{2\pi}{\lambda}(N-1)d \cos\theta}\big]^T,
\end{align}
where $\lambda$ denotes the carrier wavelength and $d$ represents the antenna spacing. 

We further define the \textit{second raw moment} as $\mathbf{R}_{k}^b =  \mathbb{E}[\mathbf{h}_{k}^b(\mathbf{h}_{k}^b)^H] \in \mathbb{C}^{N \times N},$ for the $b$-th BS and the $k$-th UE. Instead of the instantaneous channel coefficients, the second raw moment is used for localization purposes because the AoAs $\{\theta^{b,l}\}_{l=1}^{L_k^b}$ and path powers $\{\gamma^{b,l}\}_{l=1}^{L_k^b}$ evolve more smoothly with UE mobility. Note that, for simplicity, we consider only the line-of-sight (LoS) component in the subsequent framework. Next, we describe the localization principle employed in this work.
\vspace{-2mm}
\subsection{Channel charting \& Localization principle} \label{princip}
Let the $B$ BSs collect CSI samples from the $K$ UEs. The collected CSI can be classified into two types, labeled and unlabeled. We assume that $F$ labeled samples are obtained through a dedicated site survey, while the $U$ unlabeled ones are to be tracked, with $K$ given as $K = F+U$. The labeled CSI data from $b$-th BS can be expressed as $\mathcal{L}^b = \{\mathbf{R}_1^b, \cdots, \mathbf{R}_F^b\}$ with the corresponding ground-truth coordinates $\mathbf{Y} = [\mathbf{y}_1, \cdots, \mathbf{y}_F]$. Accordingly, the unlabeled CSI samples are given as $\mathcal{U}^b = \{\mathbf{R}^b_{F+1}, \cdots, \mathbf{R}^b_{F+U}\}$, and the corresponding unknown coordinates are denoted as $\widehat{\mathbf{Y}} = [\mathbf{y}_{F+1}, \cdots, \mathbf{y}_{F+U}]$. Note that, we assume multiple BSs can identify the CSI samples coming from a specific UE in the same time interval. If the pilot is not received at the $b$-th BS, the corresponding CSI $\mathbf{R}_i^b$ is set to be zero. In a static radio environment with omnidirectional antennas and fixed transmit power, the channel covariance solely depends on the UE’s spatial location. Thus, CC assumes the existence of a continuous mapping from the spatial location to the covariance-based CSI \cite{studer2018channel}. Therefore, the CSI samples $\{\{\mathbf{R}_i^b\}_{i=1}^K\}_{b=1}^B$ collected from $K$ spatial locations, denoted by $\{\mathbf{y}_{k}\}_{k=1}^K,$ described above are used to compute dimensional embeddings $\{\mathbf{z}_k\}_{k=1}^K \in \mathbb{R}^{l_d \times 1}$ via manifold learning, that preserve spatial geometry as
\begin{align}
    \parallel \mathbf{z}_u - \mathbf{z}_v \parallel_2 \: \approx \: \kappa \parallel \mathbf{y}_u -\mathbf{y}_v \parallel_2, \: \forall \: u,v \in \{1,\cdots,K\},
\end{align}
where $l_d$ represents the dimensionality of the latent space and $\kappa$ is the scaling parameter.
\captionsetup[subfloat]{labelformat=empty}
\begin{figure}
\centering
\subfloat[]{\includegraphics[scale=0.13]{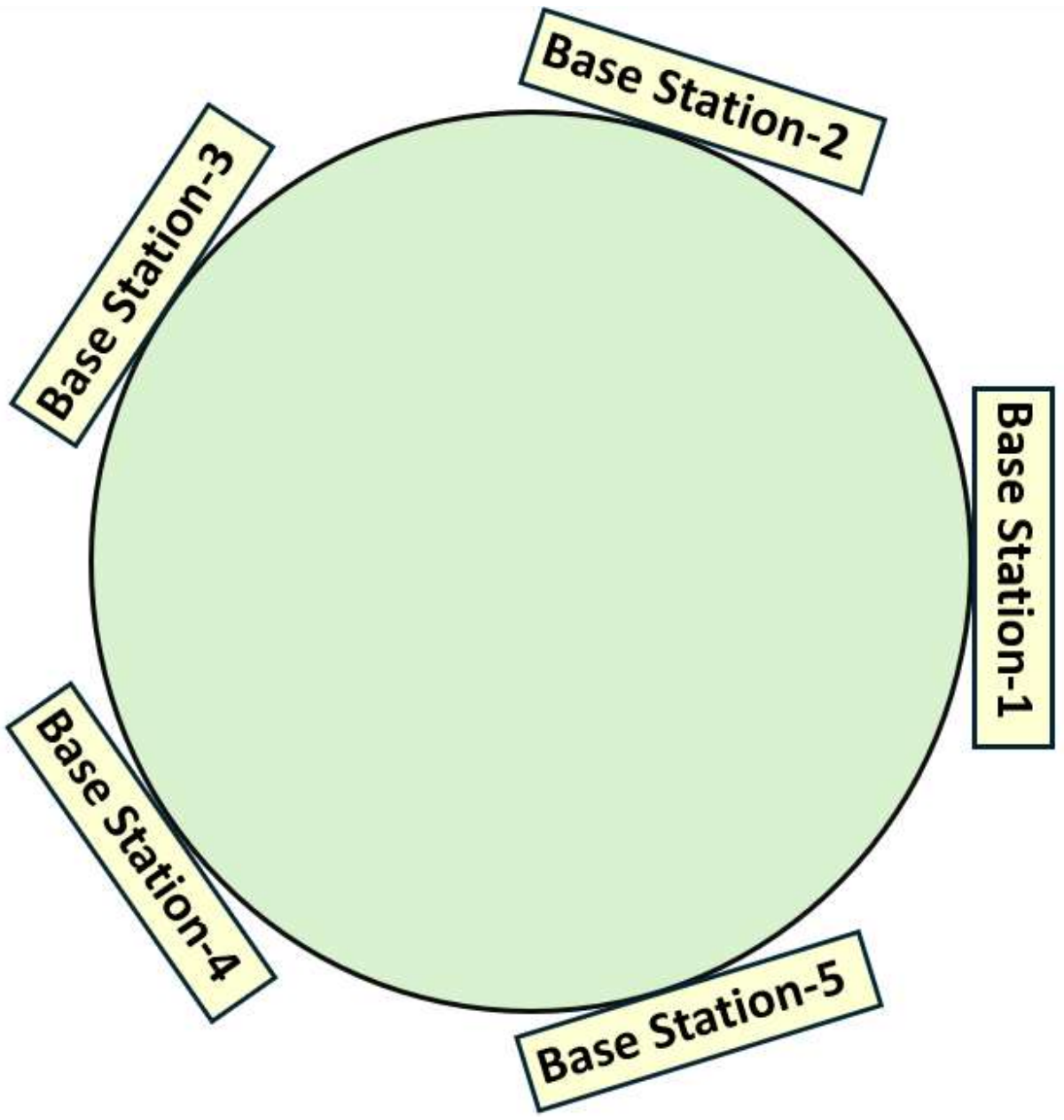}}
\hfil
\hspace{-5pt}\subfloat[]{\includegraphics[scale=0.17]{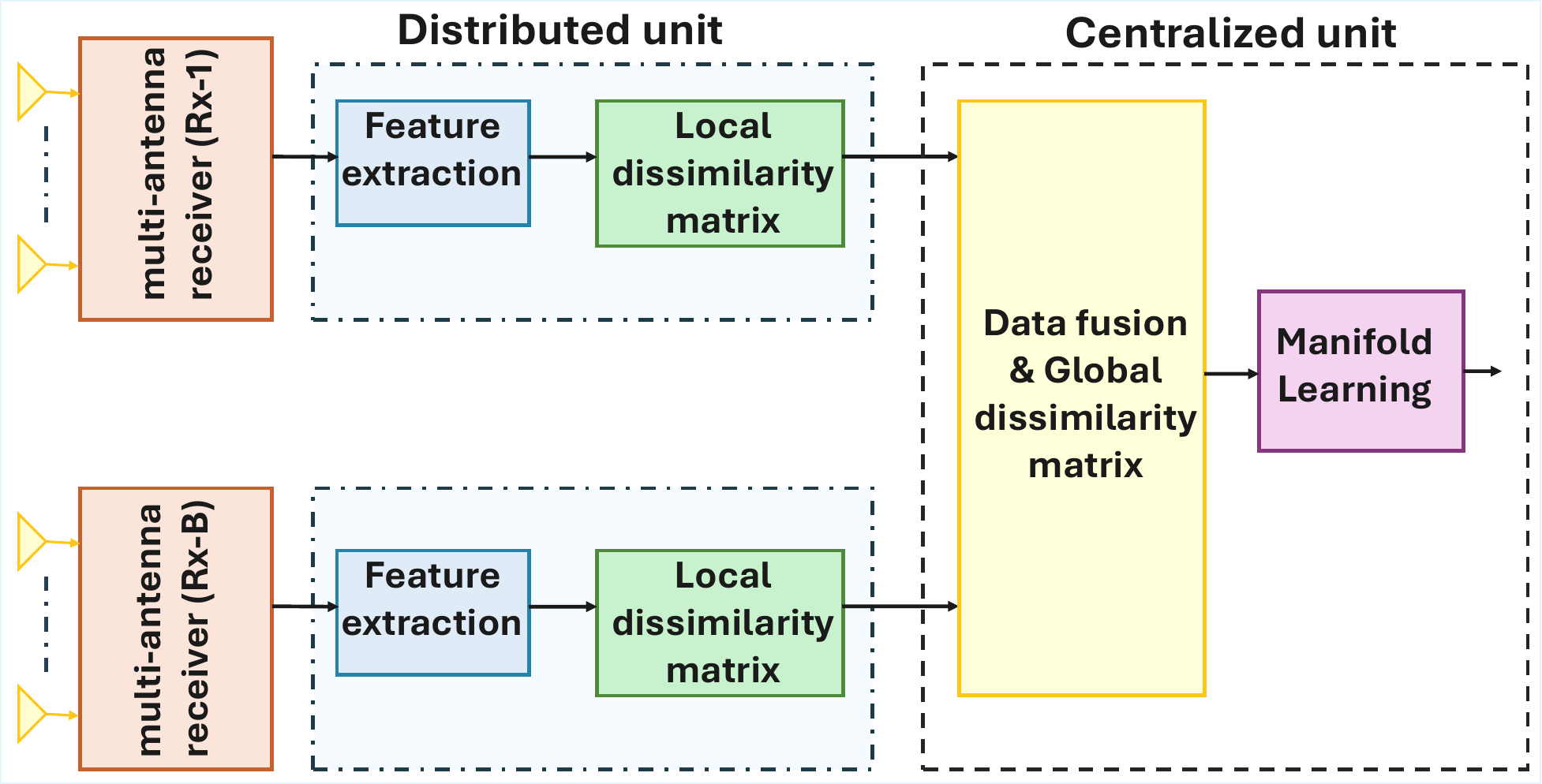}}
\vspace{-4mm}
\caption{Semi-supervised multi-point CC framework \vspace{-1 \baselineskip}}
\label{multi_CC}
\end{figure}
\vspace{-3mm}
\subsection{Feature extraction}
The CSI vector $\mathbf{h}_k^b$ by itself does not provide an effective manifold representation for CC. To obtain a meaningful geometric representation, the CSI is distilled into informative features by computing its second raw moment, scaling and transforming it into the beamspace domain, thereby emphasizing large-scale fading characteristics relevant to spatial geometry. Moreover, in order to quantify the separation between two users in the feature space, we introduce a dissimilarity measure $d_f$ (formalized in Section-\ref{simulation}) which satisfies the following property relating the second-order moments of two users (ex. A and C) to their true physical separation:
\begin{align}
    d_f(\mathbf{R}_{A}^b,\mathbf{R}_{C}^b) = |r_A^b - r_C^b|. \label{diss_measure}
\end{align}
Note that we assume users $A$, $C$ and the BS lie along a line. The quantities $r_A^b$ and $r_C^b$ are distances of users $A$ and $C$ with respect to BS.\\
\underline{\textit{CSI scaling:}} By scaling the second raw moment, one obtains
\begin{align}
    \tilde{\mathbf{R}}_{k}^b = \frac{N^{\beta-1}}{\parallel\mathbf{R}_{k}^b\parallel_\mathcal{F}^\beta}\mathbf{R}_{k}^b, \:\: \mathrm{where} \: \beta = 1+\frac{1}{2\sigma}.
\end{align}
The newly scaled moments satisfies Eq. \eqref{diss_measure} for $\sigma \in (0,\infty]$ being set as path-loss exponent of the channel.\\
\underline{\textit{Feature transform:}} The scaled moments $\tilde{\mathbf{R}}_{k}^b$ are then transformed into features by mapping them from the antenna domain to the beamspace domain to improve feature quality. This is given by $\mathbf{E}_k^b = \mathbf{Q}_{k}^b\tilde{\mathbf{R}}_{k}^b(\mathbf{Q}_{k}^b)^H$, where $\mathbf{Q}_{k}^b \in \mathbb{C}^{N \times N}$ represents the Fourier transform matrix satisfying $(\mathbf{Q}_{k}^b)^H\mathbf{Q}_{k}^b = \mathbf{I}_N$. We further vectorize the transformed matrix as $\mathrm{vec}(\mathbf{E}^b_k) = \mathbf{e}_{k}^b \in \mathbb{C}^{N^2 \times 1}$. Therefore, the vectorized features corresponding to all $K$ users at the $b$-th BS are collected and given as $\tilde{\mathbf{M}}^b = [\mathbf{e}_1^b, \cdots, \mathbf{e}_K^b] \in \mathbb{C}^{N^2 \times K}$. Let $\mathbf{M}^b = (\tilde{\mathbf{M}}^b)^T \in \mathbb{C}^{K \times N^2}$ for notational simplicity.
\vspace{-3mm}
\subsection{Local dissimilarity matrix} \label{local_diss}
At the $b$-th BS, the dissimilarity matrix $\mathbf{S}^b \in \mathbb{R}^{K \times K}$ is defined as
\vspace{-2mm}
\begin{align}
    \mathbf{S}^b = [{s}_{uv}]_{u,v=1}^K, \: \mathrm{where} \: {s}_{uv} = d_f(\mathbf{M}_u^b, \mathbf{M}_v^b).
\end{align}
The $b$-th BS collects CSI samples from its coverage area and maintains its own local view, which is captured in the dissimilarity matrix $\mathbf{S}^b$. Since each BS covers a different region of the network, they observe different subsets of the overall CSI sample set. Even for the same sample pair, the dissimilarity matrix values may differ across BSs due to varying channel conditions. As a result, one BS may hold more reliable CSI information for a given UE compared to another. This necessitates the creation of a single global dissimilarity matrix, that fuses the produced data across different multi-cell setups as shown in Fig. \ref{multi_CC}.
\vspace{-2mm}
\subsection{Global dissimilarity matrix}
As described above, we have local dissimilarity matrices $\{\mathbf{S}^b\}_{b=1}^B$. To obtain a global dissimilarity matrix $\mathbf{D} \in \mathbb{R}^{K \times K}$, we perform a weighted average of the local dissimilarity matrices collected from all $B$ BS \cite{deng2018multipoint}, which is given as
\begin{align}
    \mathbf{D} = [d_{\imath\jmath}]_{\imath,\jmath=1}^K,\: d_{\imath\jmath} = \frac{\sum_{b=1}^B w^b \mathbf{S}^b_{\imath\jmath}}{\sum_{b=1}^B w^b}.
\end{align}
The quantity $w^b = [\mathrm{min}(\tau_\imath^b,\tau_\jmath^b)]^2$ represents the weight that characterizes the reliability of dissimilarity generated by BS $b$ and $\tau_\imath^b = \mathbb{E}\{\parallel \mathbf{h}_{\imath}^b \parallel_2^2/ \sigma^2\}$ represents the signal-to-noise ratio (SNR) with $\mathbf{h}_\imath^b$ representing the channel between $b$-th BS and $\imath$-th UE, $\mathbf{y}_\imath$ denoting the UE location. Note that if a signal is not received from a particular BS $b$, we set $\mathbf{e}_{\imath}^b = \emptyset,$ $\mathbf{S}_{\imath\jmath}^b \: \mathrm{and} \:$ $\mathbf{S}_{\jmath\imath}^b, \: \forall \: \imath \neq \jmath,$ to be a large value $\psi_{\mathrm{max}},$ and the corresponding SNR $\tau_\imath^b$ is assigned a value of zero.
\vspace{-2mm}
\section{{\upshape t}-SNE to {\upshape t}-SNEkhorn: An Optimal Transport Framework} \label{SNEhorn}
t-SNE \cite{deng2021semi} is a manifold learning technique, known for its superior ability to preserve local structures when mapping high-dimensional feature vectors to a low-dimensional embedding. The framework learns a low-dimensional representation $[\mathbf{z}_1, \cdots, \mathbf{z}_K]$ by minimizing the KL divergence between the pairwise similarities of the original data and those of the low-dimensional embedding. Here, the $K$ feature vectors $\{\mathbf{e}_k\}_{k=1}^K \in \mathbb{C}^{N^2 \times 1}$ constitute the high-dimensional inputs whose structural relationships are to be retained in the learned embedding. To this end, pairwise similarities among the high-dimensional inputs are computed using Gaussian kernels, where the distances between feature vectors are obtained from the dissimilarity matrix $\mathbf{D}$. For each data point $\imath,$ a Gaussian kernel centered at $\imath$, is used to map the pairwise distances $\{d_{\imath\jmath}\}_{\jmath=1}^K$ into conditional affinities
\vspace{-2mm}
\begin{align}
    p_{\jmath|\imath} = \frac{e^{-d^2_{\jmath\imath}/ 2\sigma_\imath^2}}{\sum_{k}\:e^{-d_{k\imath}^2/2 \sigma_\imath^2}},
\end{align}
where $\sigma_\imath$ denotes the bandwidth of the Gaussian kernel. These affinities are stored in the matrix $\mathbf{P} \in \mathbb{R}^{K \times K}$. $\sigma_\imath$ is chosen such that the resulting distribution attains a perplexity, defined as $\mathrm{Perp}(\mathbf{P}_{\imath:}) = 2^{-\sum_\jmath p_{\jmath|\imath}\log_2 p_{\jmath|\imath}}$. Perplexity can be interpreted as the effective number of neighbors for each data point. It corresponds to the entropy of the distribution $\mathbf{P}_{\imath:}$ which is equivalent to that of a uniform distribution over $\mathrm{Perp}(\mathbf{P}_{\imath:})$ neighbors of $\imath$ \cite{van2023snekhorn}. Moreover, the matrix $\mathbf{P}$ possesses two key properties:
\begin{itemize}
    \item \textit{Row stochasticity:} {Each row forms a valid probability distribution $\sum_\jmath p_{\jmath|\imath} = 1, \: \forall \imath,$ ensuring that the conditional similarities are normalized across all neighbors of $\imath$.}
    \item \textit{Fixed row perplexity:} {The perplexity of each row is constrained to a user-defined value, $\mathrm{Perp}(\mathbf{P}_{\imath:}) = 2^{-\sum_\jmath p_{\jmath|\imath}\log_2 p_{\jmath|\imath}} = \phi,$ which enforces a consistent effective number of neighbors across all data points.}
\end{itemize}
Ideally, the pairwise similarity should satisfy reciprocity, i.e., $p_{\jmath|\imath} = p_{\imath|\jmath}$, which implies $\mathbf{P}_{\imath\jmath} = \mathbf{P}_{\jmath\imath}$ and thus $\mathbf{P}^T = \mathbf{P}$. Since this condition is not naturally met, a common approach is to enforce symmetry by projecting onto the space of \textit{symmetric matrices}, yielding
\begin{align}
    \mathbf{T} = \frac{\mathbf{P}+\mathbf{P}^T}{2}, \mathbf{T} = [t_{\imath\jmath}]_{\imath,\jmath=1}^K, \mathrm{where} \: t_{\imath\jmath} = \frac{p_{\jmath|\imath}+p_{\imath|\jmath}}{2}.
\end{align}
Moreover, the symmetric probability matrix $\mathbf{T} \in \mathbb{R}^{K \times K}$ is widely adopted in most t-SNE frameworks \cite{deng2018multipoint}, \cite{deng2021semi}. However, this symmetrization comes at the cost of losing two key properties originally guaranteed by $\mathbf{P}$, namely, row stochasticity and fixed row perplexity which is detailed in\footnote{Let $\mathbf{1}_K \in \mathbb{R}^{K \times 1}$ denote the all-ones vector. By construction, the matrix $\mathbf{P}$ satisfies the row-stochasticity property, i.e., $m_r(\mathbf{P}) = \mathbf{P}\mathbf{1}_K = \mathbf{1}_K.$ However, since no analogous constraint is enforced on the columns of $\mathbf{P}$, we have $m_c(\mathbf{P}) = \mathbf{P}^T \mathbf{1}_K = \mathbf{w} \in \mathbb{R}^{K \times 1}$. For the symmetrized matrix $\mathbf{T}$,
\vspace{-1mm}
\begin{align}
    m_r(\mathbf{T})=\mathbf{T} \mathbf{1}_K = \frac{\mathbf{P}\mathbf{1}_K + \mathbf{P}^T \mathbf{1}_K}{2} = \frac{\mathbf{1}_K + \mathbf{w}}{2} \neq \mathbf{1}_K, \notag
\end{align}
which clearly demonstrates that row-stochasticity is not preserved in $\mathbf{T}$. Similarly, there is a loss of the fixed-row perplexity.}. Note that, without row stochasticity and fixed row perplexity, similarity distributions become inconsistent across points, leading to {biased neighborhood modeling}. Therefore, we need a similarity matrix $\mathbf{X} \in \mathbb{R}^{K \times K}$ which satisfies the following conditions
\begin{align}
    \mathbf{X} = &\mathbf{X}^T; \sum_\jmath\mathbf{X}_{\imath\jmath} = 1; -\sum_\jmath\mathbf{X}_{\imath\jmath} \log_2 \mathbf{X}_{\imath\jmath} = \log_2\phi \: \forall \: \imath ,\notag \\ & \sum_\imath \mathbf{X}_{\imath\jmath} = 1; -\sum_{\imath}\mathbf{X}_{\imath\jmath}\log_2\mathbf{X}_{\imath\jmath} = \log_2\phi \: \forall \: \jmath. \label{matrix_prop}
\end{align}
Thus, the similarity matrix $\mathbf{X}$ is symmetric, doubly stochastic, and enforces fixed perplexity across both rows and columns. Such a formulation closely aligns with the solution structure of \textit{Entropic Optimal Transport} (OT) problems \cite{peyre2019computational}, which are typically derived through convex optimization frameworks.
\begin{algorithm}[t]
\caption{Sinkhorn-Knopp for $\mathbf{G}^*$ calculation}
\label{horn_algo}
\KwIn{Dissimilarity $\mathbf{D}$, perplexity $\phi$, scalar step sizes $\eta_u, \eta_v > 0,$ threshold $\epsilon$, max iterations $I$}
\KwOut{Optimal doubly stochastic matrix $\mathbf{G}^*$}

\textbf{Initialization:} $\mathbf{v}^{(0)} = \mathbf{1}_K, \mathbf{u}^{(0)} = \mathbf{a}^{(0)} = \mathbf{b}^{(0)} = \mathbf{0}_K, m = 0$

$\mathbf{G}^{(0)} = \mathrm{exp}^\odot \big(\frac{\mathbf{u}^{(0)}\oplus\mathbf{u}^{(0)}-2\mathbf{D}}{\mathbf{v}^{(0)} \oplus \mathbf{v}^{(0)}} \big)$

\While{$\big((m < I) \: \mathrm{and} \: (\parallel\mathbf{a}^{(m)}\parallel_\infty > \epsilon \: \mathrm{or} \parallel\mathbf{b}^{(m)}\parallel_\infty > \epsilon) \big)$}{
$\mathbf{G}^{(m)} = \mathrm{exp}^\odot \big(\frac{\mathbf{u}^{(m)}\oplus\mathbf{u}^{(m)}-2\mathbf{D}}{\mathbf{v}^{(m)} \oplus \mathbf{v}^{(m)}} \big)$

$\mathbf{a}^{(m)} = \mathbf{G} \mathbf{1}_K - \mathbf{1}_K$

$\mathbf{b}_\imath^{(m)} = \mathbf{H}(\mathbf{G}_{\imath:})- \log_2 \phi \:\: \forall \:\: \imath$ 

$\mathbf{v}^{(m+1)} = [\mathbf{v}^{(m)}-\eta_v \mathbf{b}^{(m)}]_+$

$\mathbf{u}^{(m+1)} = \mathbf{u}^{(m)} - \eta_u\mathbf{a}^{(m)}$

$m = m+1$
}
\end{algorithm}
\vspace{-3mm}
\subsection{Entropic optimal transport}\label{entropic_OT}
Entropic OT computes
\vspace{-2mm}
\begin{align}
    \textbf{P1:} \: \underset{\mathbf{G}}{\mathrm{min}} \langle \mathbf{G}, \mathbf{D} \rangle - \nu \sum_\imath \mathrm{H}(\mathbf{G}_{\imath:}),
\end{align}
over the space of doubly stochastic matrices $\{\mathbf{G} \in \mathbb{R}_+^{K \times K}: \mathbf{G}\mathbf{1}_K = \mathbf{G}^T\mathbf{1}_K = \mathbf{1}_K \}$, where $\nu > 0$ denotes the entropic regularization parameter and $\mathrm{H}(\cdot)$ represents the Shannon entropy. Considering convex duality (detailed in \cite[App.~A.2]{van2023snekhorn}), the equivalent problem is given as
\vspace{-1mm}
\begin{align}
    \textbf{P2:} \: &\underset{\mathbf{G}}{\mathrm{min}} \: \langle \mathbf{G}, \mathbf{D} \rangle \notag\\
    &\mathrm{s.t.} \: \mathbf{G}\mathbf{1}_K =\mathbf{1}_K, \mathbf{G} = \mathbf{G}^T \: \mathrm{and} \: \sum_\imath\mathrm{H}(\mathbf{G}_{\imath:}) \geq \eta, \label{P2}
\end{align}
where $0 \leq \eta \leq K\log_2K$ is a constraint on the global entropy $\sum_{\imath}\mathrm{H}(\mathbf{G}_{\imath:})$. It can be noted that the solution space in Eq. \eqref{P2} closely resembles our requirements. However, it enforces a global entropy lower bound, whereas we require a fixed row-wise perplexity constraint, as described in Eq. \eqref{matrix_prop}. Therefore, in order to impose a fixed row perplexity constraint in (P2), we consider the following
\vspace{-2mm}
\begin{align}
    \textbf{P3:} \: &\underset{\mathbf{G}}{\mathrm{min}} \langle \mathbf{G}|\mathbf{D}\rangle  \\
    & \mathrm{s.t.} \: \{\mathbf{G}\mathbf{1}_K = \mathbf{1}_K, \mathbf{G} = \mathbf{G}^T \: \mathrm{and} \: \mathrm{H}(\mathbf{G}_{\imath:}) \geq \log_2 \phi \: \forall \: \imath \}. \notag
\end{align}
Problem (P3) has a unique solution, where the entropy of each row is $\log_2 \phi$, implying that each row attains a perplexity of $\phi$ which satisfies all the required conditions as described in Eq. \eqref{matrix_prop}. The optimal solution is a \textit{unique} doubly stochastic matrix $\mathbf{G}^* \in \mathbb{R}^{K \times K}$, which is obtained by Algorithm-\ref{horn_algo}.

After obtaining the optimal similarity matrix in the high-dimensional space, we next turn to the construction of the similarity matrix in the latent space. As defined in Section-\ref{princip}, the latent-space low-dimensional coordinates are specified by $\{\mathbf{z}_k\}_{k=1}^K$. From these coordinates, we form a dissimilarity matrix $\mathbf{Q} \in \mathbb{R}^{K \times K}$, where the dissimilarity measure is considered as $\log_e(1+d_f)$. It is noteworthy that the choice of dissimilarity measure in the latent space differs from that in the input space. During manifold learning, the limited dimensionality of the latent space often leads to the \textit{crowding problem}, where distant points in the input domain are forced unnaturally close together. To mitigate this effect and better preserve the local geometry, we adopt a modified dissimilarity matrix specifically designed to balance distances in the latent representation. Moreover, unlike the high-dimensional similarity matrix, the latent-space similarity matrix does not require a fixed row-wise perplexity, since the projected coordinates are directly optimized during manifold learning, thus, (P2) is considered. The problem (P2) is solved through the Sinkhorn–Knopp algorithm, as described in Algorithm-\ref{Bstar} to obtain $\mathbf{B}^* \in \mathbb{R}^{K \times l_d}$.\\
\underline{\textit{Impact of biased neighborhood modeling on KL objective:}} By Pinsker's inequality \cite{polyanskiy2025information}, affinities $\mathbf{G}^*, \mathbf{B^*}$ satisfy
\vspace{-1mm}
\begin{align}
    \mathrm{KL}(\mathbf{G}^*,\mathbf{B}^*) \geq 2 \: (\mathrm{TV}(\mathbf{G}^*, \mathbf{B}^*))^2, \notag
\end{align}
where $\mathrm{TV}(\cdot)$ represents the total variation distance defined as $\mathrm{TV}(\mathbf{G}^*,\mathbf{B}^*) = \frac{\parallel \mathbf{G}^*-\mathbf{B}^* \parallel_1}{2} = \frac{1}{2} \sum_{i,j = 1}^K |\mathbf{G}^*_{i,j} - \mathbf{B}^*_{i,j}|$. By data processing inequality \cite{polyanskiy2025information},
\begin{align}
    \mathrm{TV}(\mathbf{G}^*,\mathbf{B}^*) \geq \mathrm{TV}(m_c(\mathbf{G}^*), m_c(\mathbf{B}^*)), \notag
\end{align}
which yields $\mathrm{KL}(\mathbf{G}^*,\mathbf{B}^*) \geq \frac{\parallel m_c(\mathbf{G}^*) - m_c(\mathbf{B}^*)\parallel_1^2}{2}$. Due to doubly stochastic nature, this lower bound is zero which is not the case for the corresponding matrices of t-SNE thereby impeding KL optimization and establishing superiority of t-SNEkhorn. The next section details the semi-supervised t-SNEkhorn framework.
\begin{algorithm}[t]
\caption{Sinkhorn-Knopp for $\mathbf{B}^*$ calculation}
\label{Bstar}
\KwIn{Dissimilarity $\mathbf{Q}$, dual variable of global entropy constraint $\chi$, threshold $\epsilon$, max iterations $I$}
\KwOut{Optimal doubly stochastic matrix $\mathbf{B}^*$}

\textbf{Initialization:} $\mathbf{u}^{(0)} = \mathbf{0}_K$, $m = 0$

$\mathbf{B} = \mathrm{exp}^\odot \big(\frac{\mathbf{u}^{(0)}\oplus\mathbf{u}^{(0)}-\mathbf{Q}}{\chi} \big)$

\While{$\big((m < I) \: \mathrm{and} \: (\parallel \mathbf{B}\mathbf{1}_K - \mathbf{1}_K \parallel_\infty > \epsilon) \big)$}{

$\mathbf{u}_\imath^{(m+1)} = \frac{1}{2}\Big(\mathbf{u}_\imath^{(m)} - \chi \log\sum_k \mathrm{exp}\Big(\frac{\mathbf{u}_k^{(m)}-\mathbf{Q}_{k\imath}}{\chi}\Big) \Big) \: \forall \:\imath$

$\mathbf{B} = \mathrm{exp}^\odot \big(\frac{\mathbf{u}^{(m+1)}\oplus\mathbf{u}^{(m+1)}-\mathbf{Q}}{\chi} \big)$

$m = m+1$
}
\end{algorithm}
\captionsetup[subfloat]{labelformat=parens}
\begin{figure*}[hbt]
\centering
\subfloat[]{\includegraphics[scale=0.32]{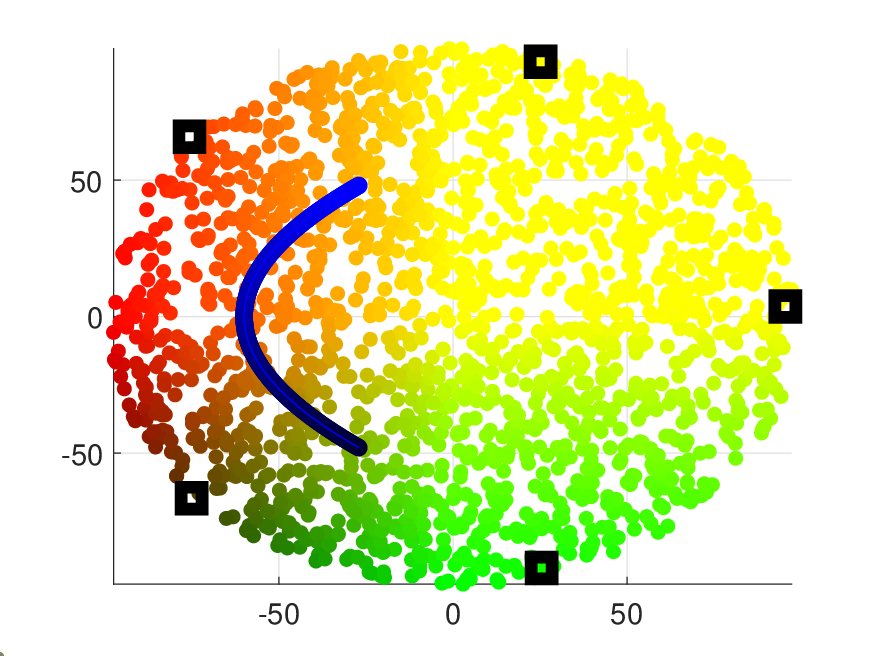}}
	\hfil
	\hspace{-15pt}\subfloat[]{\includegraphics[scale=0.32]{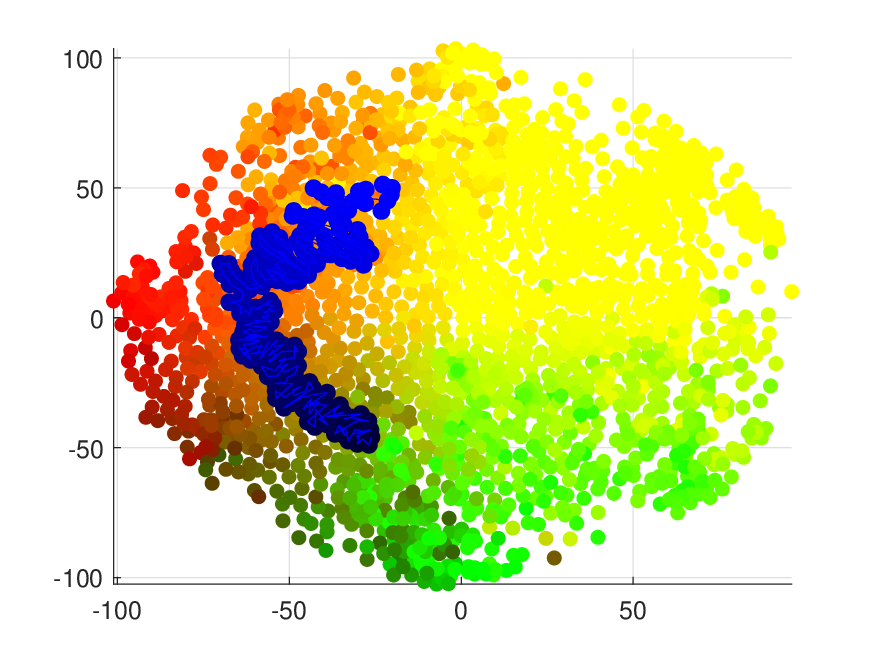}}
    \hfil
\hspace{-15pt}\subfloat[]{\includegraphics[scale=0.32]{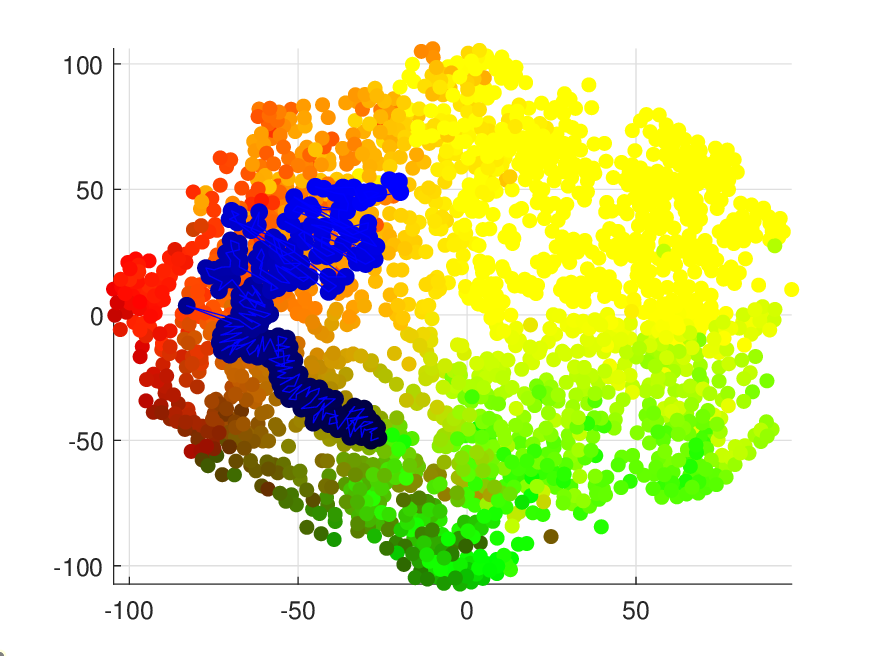}}
	\hfil
\hspace{-15pt}\subfloat[]{\includegraphics[scale=0.32]{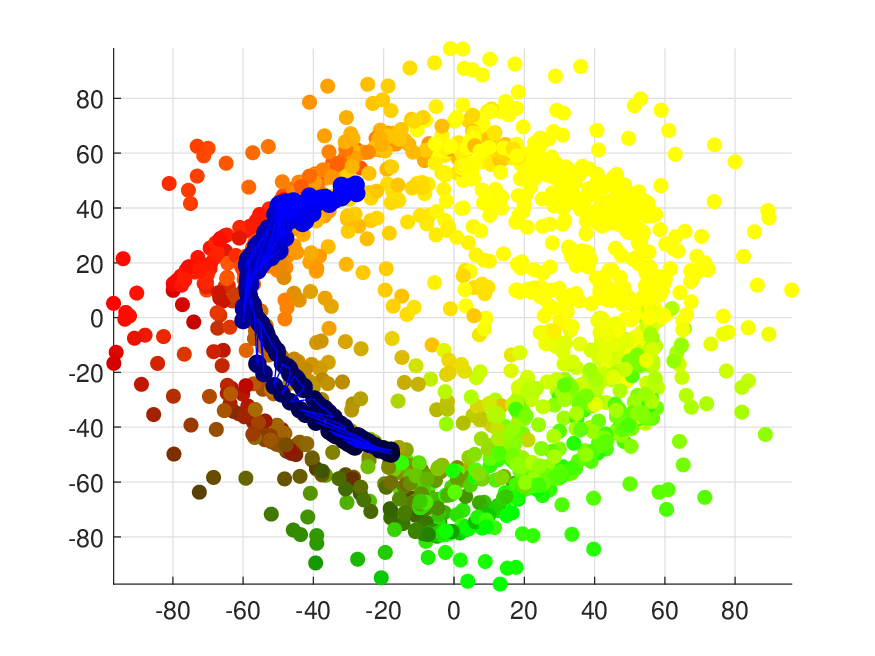}}
\vspace{-2mm}
\caption{Comparison of low-dimensional CC for different state-of-the-art algorithms $(a)$ considered scenario $(b)$ SEMIKHORN $(c)$ t-SNE $(d)$ Laplacian Eigenmaps \vspace{-1.5 \baselineskip}}
\label{scenario_}
\end{figure*}
\vspace{-2mm}
\section{SEMIKHORN: Semi-Supervised {\upshape t}-SNEKHORN}
\vspace{-0.5mm}
In line with t-SNE, the t-SNEkhorn framework utilizes KL divergence to evaluate the separation between $\mathbf{G}^*$ and $\mathbf{B}^*$, as described in Section-\ref{SNEhorn}. Therefore, the optimization problem is formulated as
\vspace{-2mm}
\begin{align}
    \textbf{P4:} \: &\underset{\mathbf{Z}}{\mathrm{min}} \: f(\mathbf{Z}) = \sum_\imath \sum_\jmath g_{\imath\jmath} \log_e\frac{g_{\imath\jmath}}{b_{\imath\jmath}}, \notag \\
    & \mathrm{s.t.} \sum_{k=1}^K \mathbf{z}_k = 0.
\end{align}
The objective function $f(\mathbf{Z})$ can be minimized using gradient descent \cite{ruder2016overview}. As the KL divergence objective is non-convex, the algorithm is typically executed with multiple random initializations and tuned perplexity values to balance global and local structure preservation, with the final embedding selected based on the minimum KL divergence. It should be noted that problem (P4) does not yield estimated UE locations. To address this, hard anchors are introduced in the learning process and the problem is re-formulated as
\vspace{-2.5mm}
\begin{align}
    \textbf{P5:} \: &\underset{\mathbf{Z}}{\mathrm{min}} \: f(\mathbf{Z}) = \sum_\imath \sum_\jmath g_{\imath\jmath} \log_e \frac{g_{\imath \jmath}}{b_{\imath \jmath}}, \notag \\
    & \mathrm{s.t.} \: \mathbf{z}_{\mathrm{\alpha}} = \mathbf{y}_{\mathrm{\alpha}}, \mathrm{\alpha} \in \mathcal{L}, \mathcal{L} = \{1,\cdots, F\}.
\end{align}
Moreover, optimizing the KL objective with labeled samples fixed at true positions generates a position map for estimating unlabeled UEs. Further, the gradient of the KL divergence between $\mathbf{G}^*$ and $\mathbf{B}^*$ is further given as
\vspace{-2mm}
\begin{align}
    \frac{\mathrm{d} \: f(\mathbf{Z})}{\mathrm{d}  \mathbf{z}_\imath} = 4 \sum_\jmath \frac{(g_{\imath\jmath}-b_{\imath\jmath})(\mathbf{z}_\imath - \mathbf{z}_\jmath)}{(1+\parallel \mathbf{z}_\imath - \mathbf{z}_\jmath \parallel^2)}. \label{div_cal}
\end{align}
The problem (P5) is solved using SEMIKHORN algorithm which is summarized in Algorithm-\ref{ss_framework}. At the end of each iteration, the coordinates of the labeled samples in the low-dimensional embedding are constrained to match their known position labels, thereby ensuring that the learned map remains aligned with the underlying geographical space. Note that the anchor values together with the perplexity parameter are optimized via \textit{Bayesian optimization}, as discussed next.
\begin{algorithm}[t]
\caption{SEMIKHORN framework}
\label{ss_framework}
\KwIn{Dissimilarity matrix $\mathbf{D},$ ground-truth coordinates $\{\mathbf{y}_1, \cdots, \mathbf{y}_F\},$ labeled set $\mathcal{L} = \{1,\cdots, F\}$}
\KwOut{Low-dimensional matrix $\mathbf{Z}^{(L)}$}

\textbf{Network Parameters:} Perplexity $\phi,$ max iterations $I$, learning rate $\eta$

\textbf{Initialization:} $\mathbf{Z}^{(0)} = \big[\mathbf{z}_1^{(0)}, \cdots, \mathbf{z}_K^{(0)}\big],$ $\mathbf{Z}^{(-1)} = \mathbf{Z}^{(0)}$

\quad $\mathbf{z}_{\mathrm{\alpha}}^{(0)} = \mathbf{y}_{\mathrm{\alpha}},$ for $\mathrm{\alpha} \in \mathcal{L},$

\quad $\mathbf{z}_{\mathrm{\alpha}}^{(0)} = \mathbf{y}_m,$ for $\mathrm{\alpha} \notin \mathcal{L}$ and $d_{\mathrm{\alpha},m}$ is smallest $\forall \: \mathrm{\alpha} \neq m$

Compute $\mathbf{G}^*$ using Algorithm-\ref{horn_algo} by solving (P3)

\For{$\ell = 1, \cdots, I$}{
Compute $\mathbf{B}^*$ using Algorithm-\ref{Bstar} by solving (P2)

Compute $\boldsymbol{\bigtriangledown} = \big[\frac{\mathrm{d} f(\mathbf{Z})}{\mathrm{d} \mathbf{z}_1}, \cdots, \frac{\mathrm{d} f(\mathbf{Z})}{\mathrm{d} \mathbf{z}_K} \big]$ using Eq. \eqref{div_cal}

Update $\mathbf{Z}^{(\ell)} = \mathbf{Z}^{(\ell-1)} + \eta \boldsymbol{\bigtriangledown} $

Set $\mathbf{z}_{\mathrm{\alpha}}^{(\ell)} = \mathbf{y}_{\mathrm{\alpha}},$ for $\mathrm{\alpha} \in \mathcal{L}$
}
\end{algorithm}
\vspace{-2mm}
\subsection{Bayesian Optimization} \label{baye_opti}
\vspace{-1mm}
To obtain the optimal value of the MLE, hyperparameters such as the similarity-matrix perplexity and the number of hard anchors are efficiently tuned using Bayesian optimization. Bayesian optimization constructs a probabilistic regression model between the hyperparameter values $\mathbf{c}_k$ and the MLE function $q(\mathbf{c}_k)$. Let $\mathbf{J}_k =\{\mathbf{c}_k,q(\mathbf{c}_k)\}_{k=1}^K$ represent the complete dataset of evaluated hyperparameters. The posterior distribution for a new sample is then given by
\vspace{-2mm}
\begin{align}
    p(q_{new}|\mathbf{c}_{new},\mathbf{J}_K) = \mathcal{N}(q_{new}|\mu(\mathbf{c}_{new}),\tilde{\sigma}^2(\mathbf{c}_{new})),
\end{align}
where $\mu(\cdot)$ and $\tilde{\sigma}^2(\cdot)$ denote the mean and variance of the predicted MLE, respectively. Let $q_{min}$ denote the minimum MLE observed so far. Therefore, the next candidate hyperparameter $\mathbf{c}_{new}$ is chosen to maximize the expected improvement (EI) criterion, defined as
\vspace{-2mm}
\begin{align}
    A_{EI}(\mathbf{c}_{new}) &= \mathbb{E}[q_{min}-q_{new}]  \\ &= \int_{-\infty}^{q_{min}}
 (q_{min}-q_{new}) p(q_{new}|\mathbf{c}_{new},\mathbf{J}_K). \notag
 \end{align}
After each iteration, the new pair $(q_{new}, \mathbf{c}_{new})$ are appended to the dataset, and the regression model is updated to guide the selection of subsequent hyperparameter values.
\captionsetup[subfloat]{labelformat=parens}
\begin{figure}[t]
\centering
\subfloat[]{\includegraphics[scale=0.26]{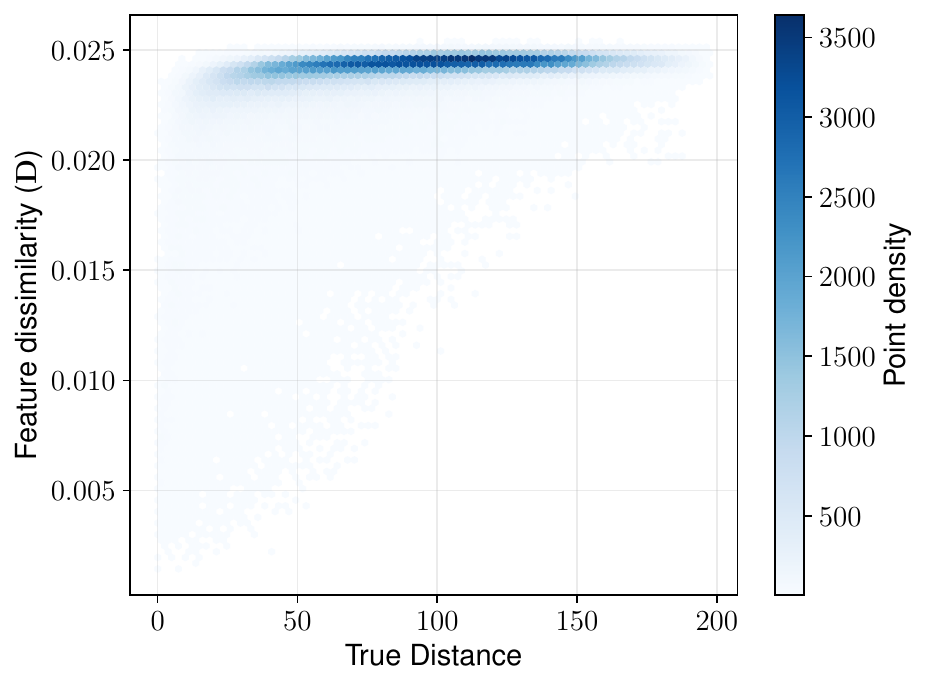}}
 \hfil
\hspace{-1pt} \subfloat[]{\includegraphics[scale=0.26]{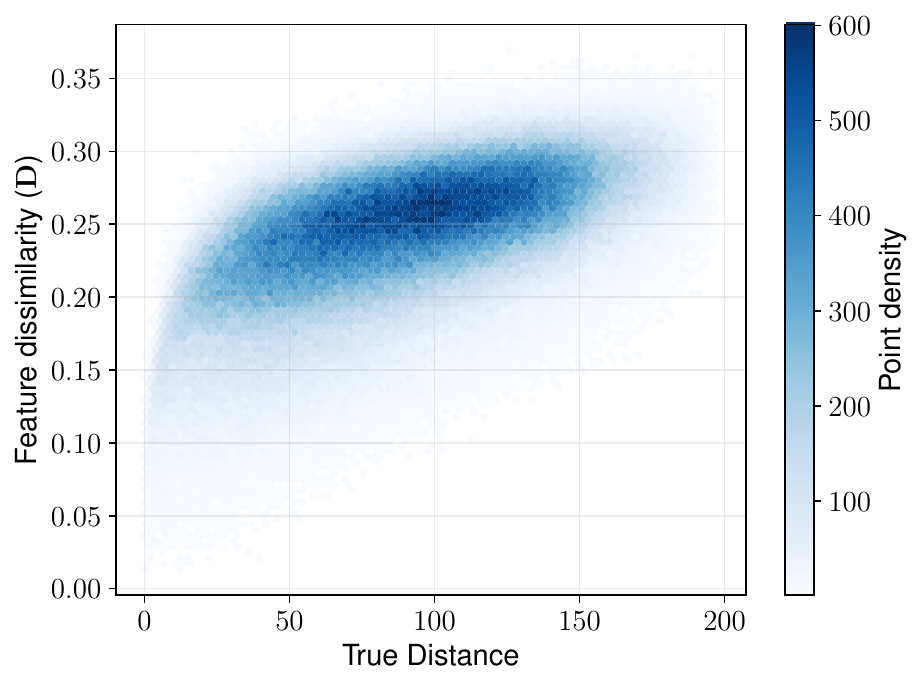}}
\vspace{-2mm}
\caption{Correlation between high dimensional space and true distance $ \left(a\right) $ Euclidean metric $ \left(b\right) $ Manhattan metric \vspace{-1.2 \baselineskip}}
\label{besl}
\end{figure}
\begin{figure*}[hbt]
\centering
\subfloat[]{\includegraphics[scale=0.31]{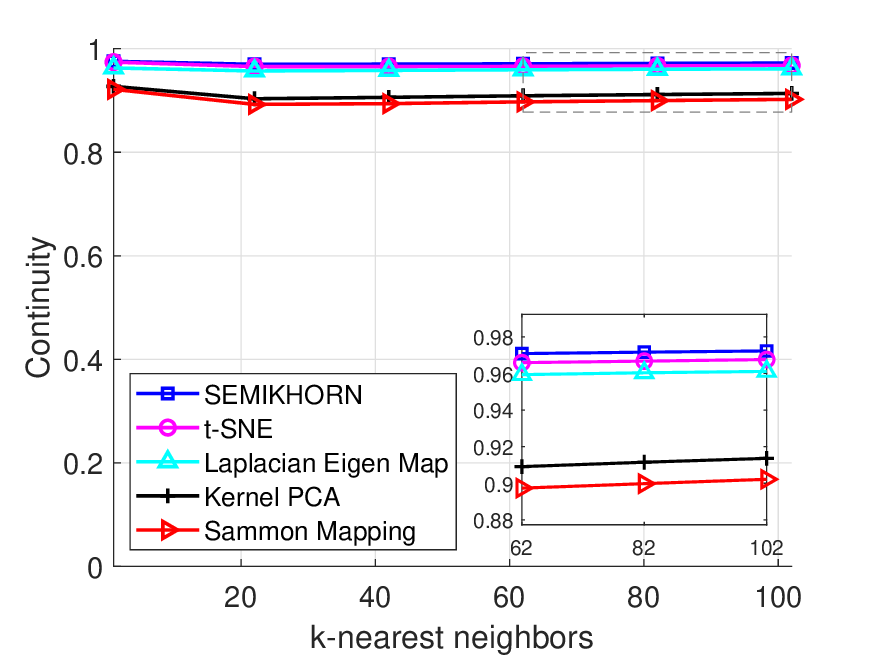}}
	\hfil
	\hspace{-20pt}\subfloat[]{\includegraphics[scale=0.31]{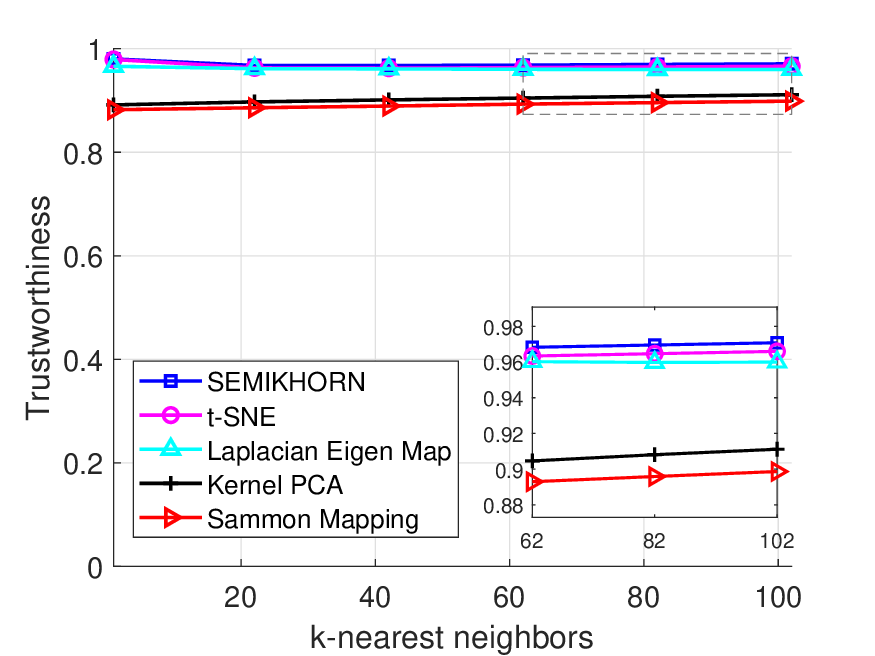}}
    \hfil
\hspace{-20pt}\subfloat[]{\includegraphics[scale=0.31]{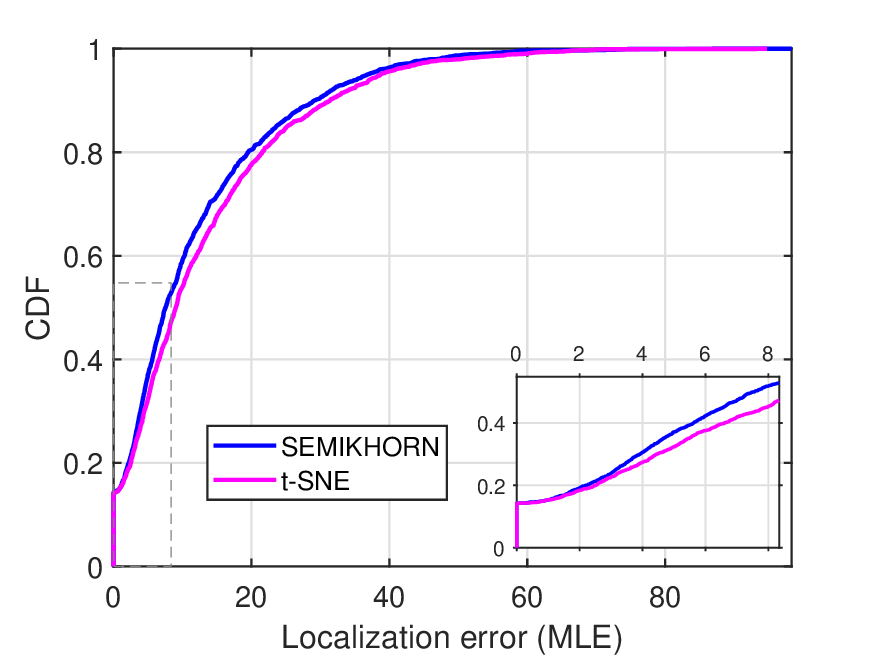}}
	\hfil
\hspace{-20pt}\subfloat[]{\includegraphics[scale=0.295]{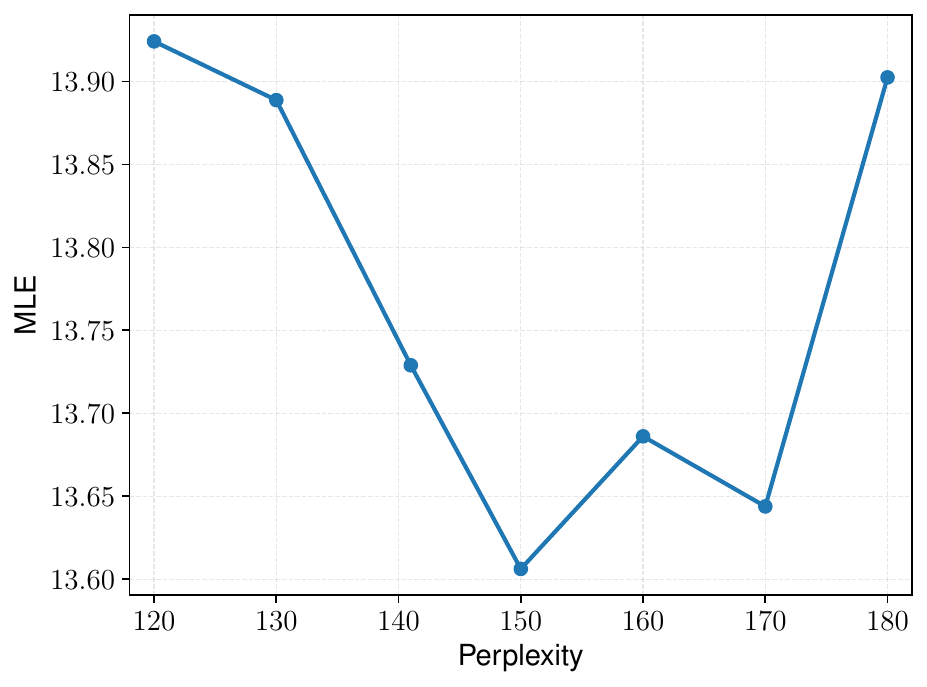}}
\vspace{-2mm}
\caption{$(a)$ Comparison of continuity for various CC algorithms $(b)$ Comparison of continuity for various CC algorithms $(c)$ CDF for localizaion errors $(d)$ Effect of perplexity in SEMIKHORN \vspace{-1.5 \baselineskip}}
\label{other_para}
\end{figure*}
\vspace{-2mm}
\section{Simulation Results} \label{simulation}
\vspace{-1mm}
We consider a dense, simulated multi-cell environment comprising $5$ mmWave BSs operating at $30$ GHz and distributed across the network, as shown in Fig. \ref{multi_CC}. Each BS is equipped with $32$ antennas arranged in a uniform linear array (ULA) with half-wavelength element spacing. CSI is collected from $2048$ UEs, which are uniformly distributed over a circular area with radius $\approx 100$m, as shown in Fig. \ref{scenario_}(a). The resulting embeddings are visualized using color gradients that encode the ground-truth spatial positions of the users, enabling assessment of global structure preservation. The blue curve in Fig. \ref{scenario_}(a) connects adjacent user positions to highlight the retention of local spatial relationships in the embedding. Further, we consider trustworthiness and continuity as the performance metrics \cite{studer2018channel}, \cite{deng2018multipoint}. \\
\underline{\textit{Manhattan distance metric:}} We consider the \textit{Manhattan distance metric} \cite{stephan2024angle} as our dissimilarity measure, which is defined by $d_f(\mathbf{A},\mathbf{B}) = \sum_{i=1}^M \sum_{j=1}^N |\mathbf{A}_{i,j}-\mathbf{B}_{i,j}|$. Fig. \ref{besl} compares the correlation of CSI features with true spatial separation for traditional Euclidean (L2 norm) \cite{studer2018channel} and Manhattan (L1 norm) distance metrics. In high-dimensional feature spaces, the Euclidean distance amplifies noise due to the squaring of component differences, which weakens the relationship between physical distance and feature dissimilarity. However, the Manhattan distance, computed through linear accumulation of absolute differences, provides a more consistent increase in dissimilarity with growing spatial separation, resulting in clearer discrimination of spatially distinct CSI samples.\\
\underline{\textit{Algorithm parameters:}} We consider $14\%$ of the total samples as labeled samples to be randomly selected. For a high-dimensional affinity matrix $\mathbf{G^*}$, we set scalar step sizes $\eta_u = \eta_v = 0.3$, tolerance $\epsilon \approx 10^{-6}$ and maximum iterations $= 4000$. For latent-space affinity matrix $\mathbf{B^*}$, we set entropic regularization $\nu = 1$, tolerance $\epsilon \approx 10^{-3}$ and maximum iterations $=30$. These parameters are chosen heuristically. Moreover, in Algorithm-\ref{ss_framework}, we consider different initialization seeds, set learning rate $\eta = 0.1$ and maximum iterations $=10,000$. The perplexity and number of hard anchors is evaluated via Bayesian optimization as described in Section-\ref{baye_opti}. 

Figs. \ref{scenario_}(b)-\ref{scenario_}(d) compare the localization results obtained from SEMIKHORN, t-SNE \cite{deng2021semi}, and Laplacian Eigenmaps \cite{deng2018multipoint}, respectively. For these channel charts, the continuity values, as depicted in Fig. \ref{other_para}(a), range between $0.95$ and $0.98$, clearly demonstrating the preservation of spatial domain neighborhood relationships in the latent space. The trustworthiness values, as depicted in Fig. \ref{other_para}(b), also lie between $0.95$ and $0.97$, indicating that most neighbors in the charts are also neighbors in the spatial domain. Moreover, SEMIKHORN achieves superior performance compared to state-of-the-art techniques due to its doubly stochastic affinity formulation, which balances pairwise relationships across all data points. This global normalization mitigates the bias of the standard t-SNE toward dense regions, resulting in improved preservation of both global topology and local neighborhood continuity compared to t-SNE and Laplacian Eigenmaps. Fig. \ref{other_para}(c) depicts the cumulative distribution function (CDF) of the absolute location errors. It can be observed that a small fraction of UE locations exhibit relatively large localization errors, which correspond to regions without nearby labeled anchor points in the semi-supervised framework. Fig. \ref{other_para}(d) shows variation of MLE with perplexity for fixed number of hard anchors obtained after Bayesian optimization. As depicted, $\phi= 150$ can preserve both the local and global manifold details and leads to MLE of $13.6$m.
\vspace{-2mm}
\section{Conclusion} \label{conclusion}
\vspace{-1mm}
In this work, we employed t-SNEkhorn-based manifold learning to embed high-dimensional CSI features onto a two-dimensional geographical map. We proposed SEMIKHORN, a semi-supervised mmWave localization framework in which multiple BSs collaboratively construct a multi-cell CC. The framework hyperparameters are further optimized using Bayesian optimization to minimize the MLE. Experimental results demonstrate that SEMIKHORN enhances the continuity and trustworthiness of the embeddings and significantly reduces MLE compared to traditional t-SNE-based CC.
\vspace{-2mm}
\section{Acknowledgment}
The authors would like to thank Saisruthi Kotamraju and Rajeev Kumar from Qualcomm, India, for their valuable suggestions and comments. The authors would like to thank the IEEE SPS Scholarship for supporting this work.
\vspace{-0.8\baselineskip}
\bibliographystyle{IEEEtran}
\bibliography{ref}

@book{polyanskiy2025information,
  title={{Information theory: From coding to learning}},
  author={Polyanskiy, Yury and Wu, Yihong},
  year={2025},
  publisher={Cambridge university press}
}

@article{stephan2024angle,
  title={{Angle-delay profile-based and timestamp-aided dissimilarity metrics for channel charting}},
  author={Stephan, Phillip and Euchner, Florian and Ten Brink, Stephan},
  journal={IEEE Transactions on Communications},
  volume={72},
  number={9},
  pages={5611--5625},
  year={2024},
  publisher={IEEE}
}

@article{van2023snekhorn,
  title={{SNEkhorn: Dimension reduction with symmetric entropic affinities}},
  author={Van Assel, Hugues and Vayer, Titouan and Flamary, R{\'e}mi and Courty, Nicolas},
  journal={Advances in Neural Information Processing Systems},
  volume={36},
  year={2023}
}

@article{ruder2016overview,
  title={{An overview of gradient descent optimization algorithms}},
  author={Ruder, Sebastian},
  journal={arXiv preprint arXiv:1609.04747},
  year={2016}
}

@inproceedings{deng2021semi,
  title={{Semi-supervised t-SNE for millimeter-wave wireless localization}},
  author={Deng, Junquan and Shi, Wei and Hu, Jian and Jiao, Xianlong},
  booktitle={2021 7th International Conference on Computer and Communications (ICCC)},
  year={2021},
  organization={IEEE}
}

@inproceedings{deng2018multipoint,
  title={{Multipoint channel charting for wireless networks}},
  author={Deng, Junquan and Medjkouh, Sa{\"\i}d and Malm, Nicolas and Tirkkonen, Olav and Studer, Christoph},
  booktitle={2018 52nd Asilomar Conference on Signals, Systems, and Computers},
  year={2018},
  organization={IEEE}
}

@article{peyre2019computational,
  title={{Computational optimal transport: With applications to data science}},
  author={Peyr{\'e}, Gabriel and Cuturi, Marco and others},
  journal={Foundations and Trends{\textregistered} in Machine Learning},
  volume={11},
  number={5-6},
  pages={355--607},
  year={2019},
  publisher={Now Publishers, Inc.}
}

@article{ferrand2023wireless,
  title={{Wireless channel charting: Theory, practice, and applications}},
  author={Ferrand, Paul and Guillaud, Maxime and Studer, Christoph and Tirkkonen, Olav},
  journal={IEEE Communications Magazine},
  volume={61},
  number={6},
  pages={124--130},
  year={2023},
  publisher={IEEE}
}

@article{studer2018channel,
  title={{Channel charting: Locating users within the radio environment using channel state information}},
  author={Studer, Christoph and Medjkouh, Sa{\"\i}d and Gonulta{\c{s}}, Emre and Goldstein, Tom and Tirkkonen, Olav},
  journal={IEEE Access},
  volume={6},
  year={2018},
  publisher={IEEE}
}

@inproceedings{wu2025improved,
  title={{An Improved Triplet-Based Channel Charting Algorithm for Positioning via Covariance Feature}},
  author={Wu, Yang and Pan, Yuhang and Ji, Jianghan and Wang, Cheng-Xiang and Li, Junling and Huang, Chen},
  booktitle={ICC 2025-IEEE International Conference on Communications},
  pages={3852--3857},
  year={2025},
  organization={IEEE}
}

@article{heath2016overview,
  title={{An overview of signal processing techniques for millimeter wave MIMO systems}},
  author={Heath, Robert W and Gonzalez-Prelcic, Nuria and Rangan, Sundeep and Roh, Wonil and Sayeed, Akbar M},
  journal={IEEE Journal of selected topics in Signal Processing},
  volume={10},
  number={3},
  pages={436--453},
  year={2016},
  publisher={IEEE}
}

@inproceedings{zhang2021semi,
  title={{Semi-supervised learning for channel charting-aided IoT localization in millimeter wave networks}},
  author={Zhang, Qianqian and Saad, Walid},
  booktitle={2021 IEEE Global Communications Conference (GLOBECOM)},
  year={2021},
  organization={IEEE}
}
\end{document}